\documentstyle[a4,11pt]{article}

\newcommand{\sect}[1]{\setcounter{equation}{0}\section{#1}\indent}
\renewcommand{\theequation}{\thesection.\arabic{equation}}
\def\appendix#1{\addtocounter{section}{1}\setcounter{equation}{0}
\renewcommand{\thesection}{\Alph{section}}
\renewcommand{\theequation}{\Alph{section}.\arabic{equation}}
\section*{Appendix \thesection\protect\indent #1}
\addcontentsline{toc}{section}{Appendix \thesection #1}}
\def\fnote#1#2{\begingroup\def\thefootnote{#1}\footnote{#2}\addtocounter
{footnote}{-1}\endgroup}

\newcommand{\win}{\hs{0}}
\font\mathbold=cmmib10 at12pt
\font\bf=cmbx10 at12pt
\newfam\bffam 
\textfont\bffam=\mathbold
\mathchardef\beta="710C
\mathchardef\gamma="710D
\mathchardef\eta="7111
\mathchardef\xi="7118
\newcommand{\BE}{\begin{equation}}
\newcommand{\EE}{\end{equation}}
\newcommand{\bea}{\begin{eqnarray}}
\newcommand{\eea}{\end{eqnarray}}
\newcommand{\ba}{\begin{array}}
\newcommand{\ea}{\end{array}}
\newcommand{\vs}[1]{\vspace{#1 mm}}
\newcommand{\hs}[1]{\hspace{#1 mm}}
\newcommand{\nonum}{\nonumber}
\def\[{{[}}
\def\]{{]}}
\def\n/{n\hs{-2.5}/}

\def\E#1#2{{e_{#1}}^{#2}} 
\def\L#1#2{{L_{#1}}^{#2}} 
\def\R#1#2{{R_{#1}}^{#2}} 
\def\gam#1#2#3#4{{({{\ga}_{#1}}^{#2})^{#3}}_{#4}} 

\renewcommand{\a}{\alpha}
\renewcommand{\b}{\beta}
\newcommand{\de}{\delta}
\newcommand{\ep}{\epsilon}
\newcommand{\la}{\lambda}

\newcommand{\ka}{\kappa}
\def\ga{\gamma}
\def\Ga{\Gamma}

\def\na{\natural}

\def\sh{\sharp}

\newcommand{\half}{\frac{1}{2}}

\def\CF{{\cal F}}

\def\S{\Sigma}

\def\phid{{\dot{\phi}}}

\def\thetad{{\dot{\theta}}}
\def\epd{{\dot{\epsilon}}}
\def\ad{{\dot{\a}}}
\def\bd{{\dot{\b}}}
\def\gad{{\dot{\ga}}}
\def\ded{{\dot{\de}}}

\def\phib{{\bar \phi}}
\def\phidb{{\bar \phid}}

\def\thetab{{\bar \theta}}
\def\epb{{\bar \ep}}

\def\thetadb{{\bar \thetad}}

\def\epdb{{\bar \epd}}

\def\phit{{\tilde \phi}}

\def\thetat{{\tilde \theta}}
\def\thetatb{{\bar \thetat}}
\def\ept{{\tilde \epsilon}}
\def\eptb{{\bar \ept}}

\def\at{{\tilde \a}}
\def\bt{{\tilde \b}}
\def\gat{{\tilde \ga}}
\def\det{{\tilde \de}}

\def\phitb{{\bar \phit}}

\def\ra{\partial}

\font\mathbold=cmmib10 at12pt
\font\bf=cmbx10 at12pt
\newfam\bffam 
\textfont\bffam=\mathbold
\mathchardef\beta="710C
\mathchardef\gamma="710D
\mathchardef\eta="7111
\mathchardef\xi="7118
\begin{document}
\topmargin -38pt
\oddsidemargin -6mm
\newcommand{\NP}[1]{Nucl.\ Phys.\ {\bf #1}}
\newcommand{\PL}[1]{Phys.\ Lett.\ {\bf #1}}
\newcommand{\CMP}[1]{Comm.\ Math.\ Phys.\ {\bf #1}}
\newcommand{\PR}[1]{Phys.\ Rev.\ {\bf #1}}
\newcommand{\PRL}[1]{Phys.\ Rev.\ Lett.\ {\bf #1}}
\newcommand{\PTP}[1]{Prog.\ Theor.\ Phys.\ {\bf #1}}
\newcommand{\PTPS}[1]{Prog.\ Theor.\ Phys.\ Suppl.\ {\bf #1}}
\newcommand{\MPL}[1]{Mod.\ Phys.\ Lett.\ {\bf #1}}
\newcommand{\IJMP}[1]{Int.\ Jour.\ Mod.\ Phys.\ {\bf #1}}
\begin{titlepage}
\setcounter{page}{0}
\begin{flushright}
KEK-TH-590\\
KEK Preprint 98-152\\
hep-th/9809113\\
September, 1998\\
\end{flushright}
\vs{0}
\begin{center}
\vs{20}
{\bf {\Large Type-II Superstrings and New Spacetime Superalgebras 
}}\\
\vs{30}

{\large Makoto SAKAGUCHI\fnote{\star}{e-mail:
gu@theory.kek.jp}}\\
\vs{7}
{\em
Theory Division, Institute of Particle and Nuclear Studies,\\
High Energy Accelerator Research Organization (KEK),\\
1-1 Oho, Tsukuba, Ibaraki, 305-0801
Japan}\\
\end{center}
\vs{25}

\baselineskip 16pt
\centerline{\bf Abstract}
\vs{8}
\win
We present a geometric formulation of type-IIA and -IIB superstring theories
in which the Wess-Zumino term is second order in the supersymmetric currents.
The currents are constructed using supergroup manifolds corresponding to
superalgebras: the IIA superalgebra derived from M-algebra
and the IIB superalgebra obtained by a T-duality transformation of the IIA
superalgebra.
We find that a slight modification of the IIB superalgebra is needed
to describe D-string theories,
in which the U(1) gauge field on the worldsheet is explicitly constructed
in terms of D-string charges.
A unification of the superalgebras in a $(10+1)$-dimensional $N=2$ superalgebra
is discussed too.

\end{titlepage}
\newpage
\baselineskip 18pt
\newpage
\renewcommand{\thefootnote}{\arabic{footnote}}
\setcounter{footnote}{0}
\sect{Introduction}
\win
It is now widely appreciated that
super $p$-branes play an important role in non-perturbative string physics.
The dynamics of the super $p$-branes is generally awfully difficult,
but it does possess some algebraic properties.
One of these is a modification of the Poincar\'e superalgebra
in the presence of super $p$-branes.

\win
Siegel~\cite{Siegel} found a manifestly supersymmetric
formulation of the Green-Schwarz superstring,
based on a superalgebra discovered earlier by Green~\cite{Green}.
This superalgebra is a generalization of super Poincar\'e algebra,
in which a new fermionic generator is contained and translations do not commute with the
supercharges.
He constructed a suitable set of supercurrents on the corresponding supergroup manifold
and wrote down the Wess-Zumino term of the Green-Schwarz action in a manifestly
supersymmetric form, without having go to one higher dimension.
Bergshoeff and Sezgin showed that Siegel's formulation generalizes
to higher super $p$-branes~\cite{BS}.
They introduced a set of new spacetime superalgebras.
By introducing the new coordinates corresponding to the new generators
of the underlying superalgebra
and constructing the supercurrents on the supergroup manifolds,
they were able to write the Wess-Zumino terms for super $p$-branes,
which are $(p+1)$-th order in the supercurrents and
which equal the usual Wess-Zumino terms up to the total derivative terms.

\win
It is not known whether the formulation generalizes to
type-II branes:
type-II superstrings, NS5-branes and
D $p$-branes ($p=$odd for the IIB superstring theory
and $p=$even for the IIA superstring theory).
In this paper, we show that the formulation generalizes
to type-II superstrings and D-strings.
To do this,
we introduce a set of new spacetime superalgebras:
the IIA superalgebra derived from the M-algebra
which was discovered by Sezgin~\cite{M-algebra}
and the IIB superalgebra obtained by a T-duality transformation of the IIA
superalgebra.
Using the new superalgebras,
we construct supercurrents on the supergroup manifolds
corresponding to the superalgebras.
In terms of these currents,
we write down the Wess-Zumino terms,
which are second order in the supercurrents.
We find that one needs a slight modification of the IIB superalgebra
in order to describe D-string theories,
in which the U(1) gauge field on the worldsheet is parametrized by coordinates
associated with the D-string charges.
The modified IIB superalgebra,
which corresponds to the description of the IIB superstring and the D-string on an equal footing,
is not related by the T-duality to the IIA superalgebra obtained from the M-algebra.
As a trial to relate these superalgebras,
we consider a unification of the superalgebras in a $(10+1)$-dimensional $N=2$ superalgebra.

\win
This paper is organized as follows.
We first present a review of the technology used in this paper
and Siegel's formulation.
In sec.3, we derive the IIA superalgebra from the M-algebra
and show that the superalgebra corresponds to the IIA superstring theories.
Next, in sec.4,  performing a T-duality transformation,
we obtain the IIB superalgebra.
The algebra is shown to correspond to the IIB superstrings.
In sec.5, D-strings are found to be described by modifying the IIB superalgebra,
in which the U(1) gauge field is parametrized by the coordinates corresponding to
the D-string charges.
In sec.6, considering the identities, we show that
the modified IIB superalgebra can not be related to the IIA superalgebra
by T-duality transformations.
A unification of these algebras in a $(10+1)$-dimensional $N=2$ superalgebra
is discussed in sec.7.
The last section is devoted to a summary and discussions.

\sect{Superalgebra and Siegel's Formulation}\label{}
\win
Let us denote the generators of an algebra collectively as $T_A$.
The algebra can be written as
\bea
\[T_A,T_B\}={f_{AB}}^CT_C.
\eea
In the dual basis, the Maurer-Cartan super 1-forms are defined by
\bea
e^A=dZ^M{L_{M}}^A,
\eea
where $dZ^M$ are the differentials on the group manifold.
The left-invariant group vielbeins ${L_M}^{A}$ and the pullbacks ${L_i}^{A}$ are
obtained by the left-invariant Maurer-Cartan form,
\bea
U^{-1}\ra_iU=\ra_iZ^M\L{M}{A} T_A=\L{i}{A}T_A,
\eea
where $U$ is a supergroup element.
The Maurer-Cartan equations, which are expressed in terms of the dual forms,
\bea
de^C=-\half e^B\wedge e^A {f_{AB}}^C,
\eea
contain equivalent informatin about the algebra.
Given a super $q$-form $G$, the exterior derivative acts as follows:
$d(F\wedge G)=F\wedge dG+(-1)^qdF\wedge G$.
The Jacobi identities are satisfied
iff the integrability conditions, $d^2 e^A=0$, hold.

\win
Similarly, the right-invariant group vielbeins ${R_M}^{A}$
are obtained by the right-invariant Maurer-Cartan form,
\bea
\ra_iU U^{-1}=\ra_iZ^M \R{M}{A} T_A=\R{i}{A}T_A.
\label{right}
\eea
Using the the right-invariant group vielbeins $\R{M}{A}$,
supersymmetry transformations are obtained as follows.
An infinitesimal transformation is written as
$U'=(1+\ep)U$, where $\ep=\ep^AT_A$ is the transformation parameter.
This implies that $\ep =\de UU^{-1}=\de Z^B \R{B}{A}T_A$,
one then finds that an infinitesimal transformation can be expressed as
\bea
\de Z^A = \ep^B \hat R_B{}^{A},
\eea
where $ \hat R_B{}^A$ is defined by $\R{M}{A} \hat R_A{}^N=\de^{N}_{M}$.
The transformation parameter $\ep^\a$ associated with the supercharge $Q_\a$
can be interpreted as a rigid
supersymmetry transformation parameter.

\win
Useful in calculating the left-/right-invariant Maurer-Cartan equations are 
the Zumino's formulae:
\bea
e^{-\phi}de^{\phi}  &=& ( \frac{1-e^{-\phi}}{\phi} ) \wedge d\phi ,\\
de^{\phi}e^{-\phi} &=& ( \frac{e^{\phi}-1}{\phi} )\wedge d\phi ,\\
e^{-\phi}\b e^{\phi} &=& e^{-\phi}\wedge \b,\\
e^{\phi}\b e^{-\phi} &=& e^{\phi}\wedge \b,
\eea
where the wedge denotes a compact expression of the commutation relatioln:
$\phi\wedge\psi=\[\phi,\psi\]$, $\phi^2\wedge\psi=\[ \phi,\[\phi,\psi\]\]$ and
$1\wedge\psi=\psi$.

\win
Siegel~\cite{Siegel} found a manifestly supersymmetric formulation of the Green-Schwarz
superstring based on a superalgebra~\cite{Green},
\bea
\{Q_\a,Q_\b \}= \gam{}{a}{}{\a\b}P_a,~~~
\[Q_\a,P_a\]=\gam{a}{}{}{\a\b}\Sigma^\b,
\eea
in which the translation does not commute with the supercharge.
Parametrizing the supergroup manifold as
\BE
U=e^{\phi_\a \S^\a} e^{x_a P^a} e^{\theta^\a Q_\a},
\EE
where coordinates $Z^A=(\phi_\a, x_a, \theta^\a)$
associate to the generators $T_A=(\S^\a, P^a, Q_\a)$,
the superstring action
is written in terms of left-invariant pullback vielbeins,
\bea
I=\int d^2\xi \[ -\half\sqrt{-g}g^{ij}\L{i}{a} \L{j}{b}\eta_{ab}
                 -\half \ep^{ij}\L{i}{\a} \L{j\a}{} \],
\eea
where $g_{ij}$ and $\eta_{ab}$ are the worldsheet and spacetime metric, respectively
and $g= {\rm det} g_{ij}$.
A nontrivial feature of this new action is that the new coordinates
$\phi_\a$ only occur as a total derivative term.
Up to this total derivative term the above action is identical to the
standard Green-Schwarz superstring action.
Furthermore, the Wess-Zumino term in the above action
is manifestly supersymmetric,
while in the usual Green-Schwarz formulation the supersymmetry
is up to a total derivative term.
These transformations involve $\L{i}{a}$ and $\L{i}{\a}$, which remain unchanged
by the presence of the new coordinate $\phi_\a$.
The coefficient of Wess-Zumino term is fixed so that
the action is also invariant under the usual $\kappa$-symmetry transformations.
Following the same line, the authors of ref.\cite{BS} showed that
$p$-brane actions can be constructed by using new superalgebras.

\sect{IIA Superstring}
\win
We derive the IIA superalgebra by a dimensional reduction of the M-algebra.
For later use in sec.\ref{IIB},
we include D0-brane and D2-brane charges in addition to
supertranslation and superstring charges.
We then show that the IIA superstring action can be constructed
using the obtained IIA superalgebra.

\subsection{IIA superalgebra from M-algebra}
\win
The M-algebra found by Sezgin~\cite{M-algebra} is characterized by generators:
supertranslation $Q_M$, M2-brane $Z^{MN}$,
``superstring''\footnote{
This was discussed in \cite{Curtlight} and \cite{Sezgin}
}
$Z^M$ and M5-brane $Z^{MNOPQ}$ ,
where 11-dimensional spacetime indices,
$\mu,\nu,\cdots$
and Majorana spinor indices, $\a,\b,\cdots$
are collectively denoted as $M,N,\cdots$,
so that $Q_M=(P_\mu,Q_\a)$, $Z^{MN}=(Z^{\mu\nu},Z^{\mu\a},Z^{\a\b})$ etc.
The Maurer-Cartan equations, containing equivalent information about the algebra,
are described in terms of the dual basis: $e^M$, $e_{MN}$, $e_M$ 
and $e_{MNOPQ}$, respectively.
It is sufficient for our present purpose to consider the former three:
$e^M$, $e_{MN}$ and $e_M$.
The corresponding part of the M-algebra is as follows
(we call this algebra M-algebra for simplicity throughout this paper):
\bea
de^\mu&=&-\half e^\a\wedge e^\b(\ga^\mu)_{\a\b},\\
d{e_\mu}&=&-\half e^\a\wedge e^\b(\ga_\mu)_{\a\b},\\
d{e_\a}&=& -e^\b\wedge e^\mu(\ga_\mu)_{\a\b}
            +(1-\la)e^\b\wedge {e_\mu}'(\ga^\mu)_{\a\b}
            -\frac{\la}{10}e^\b \wedge e_{\mu\nu}(\ga^{\mu\nu})_{\a\b},
\label{3inMalgebra}\\
de_{\mu\nu}&=& -\half e^\a\wedge e^\b(\ga_{\mu\nu})_{\a\b},
\label{4inMalgebra}\\
de_{\mu\a}&=& - e^\b\wedge e^\nu(\ga_{\mu\nu})_{\a\b}
              - e^\b\wedge e_{\mu\nu}(\ga^{\nu})_{\a\b},
\label{5inMalgebra}\\
de_{\a\b}&=& \half e^\mu\wedge e^\nu(\ga_{\mu\nu})_{\a\b}
            -\half e_{\mu\nu}\wedge e^\mu(\ga^{\nu})_{\a\b}
            -\frac{1}{4} e_{\mu\ga} \wedge e^\ga(\ga^{\mu})_{\a\b}
            -2 e_{\mu\a}\wedge e^\ga(\ga^{\mu})_{\b\ga},
\label{6inMalgebra}
\eea
where the Jacobi identities are satisfied
due to the identities in $(10+1)$-dimensions:
\bea
&&(\ga_\mu)_{(\a\b}(\ga^{\mu\nu})_{\ga\de)}=0,
\label{identityM1}\\
&&(\ga_\mu)_{(\a\b}(\ga^{\mu})_{\ga\de)}
+\frac{1}{10} (\ga_{\mu\nu})_{(\a\b}(\ga^{\mu\nu})_{\ga\de)}=0.
\label{identityM2}
\eea
Throughout this paper we use a notation where a given spinor always has an
upper or a lower spinor-index,
and never raise or lower a spinor
index using the charge-conjugation matrix.

\win
The IIA superalgebra is obtained by a dimensional reduction
of the 11-th direction, say $x^\na$.
The obtained IIA superalgebra is characterized by generators:
supertranslation $Q_A$,
superstring $Z^A$,
D0-brane $\S$,
D2-brane $\S^{AB}$
and ${Z^A}'$ and ${Z^\na}'$ originated from ``superstring'' $Z^M$ in 11-dimensions,
where the indices $A,B,\cdots$ collectively denote 10-dimensional spacetime indices,
$a,b,\cdots$
and Majorana-Weyl spinor indices, $\a,\b,\cdots$ and $\ad,\bd,\cdots$
with positive and negative chirality, respectively.
We find that the dual forms of the generators of the IIA superalgebra are defined
in terms of those of the generators of the M-algebra, as follows:
\bea
\begin{tabular}{rrl}
{\rm Supertranslation $Q_A$:}& $e^A$&$=(e^a,e^\a,e^\ad)$,\\
{\rm Superstring $Z^A$:}&$e_A$&$=(e_{\na a},e_{\na \a},e_{\na \ad})$,\\
{\rm D0-brane $\S$:}&$e$&$=(e^\na)$,\\
{\rm D2-brane $\S^{AB}$:}&$e_{AB}$&$=(e_{ab}, e_{a\a}, e_{a\ad}, e_{\a\b}, e_{\a\bd},
e_{\ad\bd})$,\\
${Z^A}'$: &${e_A}'$&$=({e_a}', {e_\a}', {e_\ad}')$,\\
${Z^\na}':$&$e'$&$=({e_\na}')$,
\end{tabular}
\eea
which is consistent with the fact
that superstring and D2-brane consist of wrapped M2-brane and M2-brane,
respectively.
The Maurer-Cartan equations of the IIA superalgebra are found to be
\bea
d\E{}{a}&=&-\half \E{}{\a}\wedge \E{}{\b} \gam{}{a}{}{\a\b}
           -\half \E{}{\ad}\wedge \E{}{\bd} \gam{}{a}{}{\ad\bd}\label{IIAMC1},\\
d\E{a}{}&=&-\half \E{}{\a}\wedge \E{}{\b} \gam{a}{}{}{\a\b}
           +\half \E{}{\ad}\wedge \E{}{\bd} \gam{a}{}{}{\ad\bd}\label{IIAMC2},\\
d\E{\a}{}&=&- \E{}{\b}\wedge \E{}{a} \gam{a}{}{}{\a\b}
            -\E{}{\b}\wedge \E{a}{} \gam{}{a}{}{\a\b},\\
d\E{\ad}{}&=& +\E{}{\bd}\wedge \E{}{a} \gam{a}{}{}{\ad\bd}
            -\E{}{\b}\wedge \E{a}{} \gam{}{a}{}{\ad\bd},\\
d\E{}{}&=&-\E{}{\a}\wedge \E{}{\bd} (1)_{\a\bd},\\
d\E{ab}{}&=&- \E{}{\a}\wedge \E{}{\bd} \gam{ab}{}{}{\a\bd},\\
d\E{a\a}{}&=&-\E{}{\bd}\wedge \E{}{b} \gam{ab}{}{}{\a\bd}
             +\E{}{\b}\wedge \E{}{} \gam{a}{}{}{\a\b}
             -\E{}{\b}\wedge \E{ab}{} \gam{}{b}{}{\a\b}
             +\E{}{\bd}\wedge \E{a}{} (1)_{\a\bd},\\
d\E{a\ad}{}&=&-\E{}{\b}\wedge \E{}{b} \gam{ab}{}{}{\ad\b}
             -\E{}{\bd}\wedge \E{}{} \gam{a}{}{}{\ad\bd}
             -\E{}{\bd}\wedge \E{ab}{} \gam{}{b}{}{\ad\bd}
             -\E{}{\b}\wedge \E{a}{} (1)_{\ad\b},\\
d\E{\a\b}{}&=&+\E{}{}\wedge \E{}{a} \gam{a}{}{}{\a\b}
             -\half \E{ab}{}\wedge \E{}{a} \gam{}{b}{}{\a\b}
             -\half \E{b}{}\wedge \E{}{} \gam{}{b}{}{\a\b}
             -\frac{1}{4} \E{a\ga}{}\wedge \E{}{\ga} \gam{}{a}{}{\a\b}~~~\nonum\\
    &&       -\frac{1}{4} \E{a\gad}{}\wedge \E{}{\gad} \gam{}{a}{}{\a\b}
             -2 \E{a\a}{}\wedge \E{}{\ga} \gam{}{a}{}{\b\ga}
             -2 \E{\a}{}\wedge \E{}{\gad} (1)_{\b\gad},\\
d\E{\a\bd}{}&=&+\half \E{}{a}\wedge \E{}{b} \gam{ab}{}{}{\a\bd}
              +\half \E{a}{}\wedge \E{}{a} (1)_{\a\bd}
             -\frac{1}{4} \E{\ga}{}\wedge \E{}{\ga} (1)_{\a\bd}
             -\frac{1}{4} \E{\gad}{}\wedge \E{}{\gad} (1)_{\a\bd}\nonum\\
      &&     - \E{a\a}{}\wedge \E{}{\gad} \gam{}{a}{}{\bd\gad}
             +\E{\a}{}\wedge \E{}{\ga} (1)_{\bd\ga}
            - \E{a\bd}{}\wedge \E{}{\ga} \gam{}{a}{}{\a\ga}
             - \E{\bd}{}\wedge \E{}{\gad} (1)_{\a\gad},\\
d\E{\ad\bd}{}&=&-\E{}{}\wedge \E{}{a} \gam{a}{}{}{\ad\bd}
             -\half \E{ab}{}\wedge \E{}{a} \gam{}{b}{}{\ad\bd}
             -\half \E{b}{}\wedge \E{}{} \gam{}{b}{}{\ad\bd}
             -\frac{1}{4} \E{a\ga}{}\wedge \E{}{\ga} \gam{}{a}{}{\ad\bd}~~~\nonum\\
    &&       -\frac{1}{4} \E{a\gad}{}\wedge \E{}{\gad} \gam{}{a}{}{\ad\bd}
             -2 \E{a\ad}{}\wedge \E{}{\gad} \gam{}{a}{}{\bd\gad}
             +2 \E{\ad}{}\wedge \E{}{\ga} (1)_{\bd\ga},\\
d{\E{a}{}}'&=&-\half e^\a\wedge e^\b(\ga_a)_{\a\b}
              -\half e^\ad\wedge e^\bd(\ga_a)_{\ad\bd}\label{primed1},\\
d{\E{\a}{}}'&=& -e^\b\wedge e^a (\ga_a)_{\a\b}
                -e^\bd\wedge e (1)_{\a\bd}
                +(1-\la)e^\b\wedge {e_a}' (\ga^a)_{\a\b}\nonum\\
            &&  +(1-\la)e^\bd\wedge e' (1)_{\a\bd}
                -\frac{\la}{10}e^\bd\wedge e_{ab} (\ga^{ab})_{\a\bd}
                -\frac{\la}{5}e^\b\wedge e_a (\ga^a)_{\a\b},
\label{primed2}\\
d{\E{\ad}{}}'&=& -e^\bd\wedge e^a (\ga_a)_{\ad\bd}
                +e^\b\wedge e (1)_{\ad\b}
                +(1-\la)e^\bd\wedge {e_a}' (\ga^a)_{\ad\bd}\nonum\\
            &&  -(1-\la)e^\bd\wedge e' (1)_{\ad\b}
                -\frac{\la}{10}e^\b\wedge e_{ab} (\ga^{ab})_{\ad\b}
                +\frac{\la}{5}e^\bd\wedge e_a (\ga^a)_{\ad\bd},
\label{primed3}\\
d{\E{}{}}'&=& - e^\a \wedge e^\bd(1)_{\a\bd},\label{primed4}
\eea
where we used the following relations~\cite{KT}: $(1)_{\a\bd}=-(1)_{\bd\a}$
and $\gam{ab}{}{}{\a\bd}=+\gam{ab}{}{}{\bd\a}$.
The IIA superalgebra is closed due to the following identies:
\bea
&&(\ga_a)_{\a(\b}(\ga^a)_{\ga\de)}=0,\label{identityIIA1}\\
&&(1)_{\ad(\b}(\ga^a)_{\ga\de)}+(\ga^a_{~b})_{\ad(\b}(\ga^b)_{\ga\de)}=0,\label{identityIIA2}\\
&&-(1)_{\a(\bd}(1)_{\gad)\de}+\frac{2}{5}(\ga_a)_{\a\de}(\ga^a)_{(\bd\gad)}
+\frac{1}{10}(\ga_{ab})_{\a(\bd}(\ga^{ab})_{\gad)\de}=0,
\label{identityIIA3}
\eea
and those in which the undotted spinor indices
are exchanged for the dotted ones.
The last identity (\ref{identityIIA3}) is needed to satisfy
the Jacobi identity for primed dual forms
originating from ``superstring'' in the M-algebra.

\subsection{IIA superstring}
\win
We now consider a subalgebra of the IIA superalgebra
generated by the following generators:
$T_A$ $=$ $\{P_a$, $Q_\a$, $Q_\ad$, $Z^a$, $Z^\a$, $Z^\ad \}$.
The algebra turns out to be
\bea
\{ Q_\a ,Q_\b\}&=&\gam{}{a}{}{\a\b}P_a+\gam{a}{}{}{\a\b}Z^a,\nonum\\
\{ Q_\ad ,Q_\bd\}&=&\gam{}{a}{}{\ad\bd}P_a-\gam{a}{}{}{\ad\bd}Z^a,\nonum\\
\[P_b,Q_\b\]&=&2\gam{b}{}{}{\a\b}Z^\a,~~~
\[Z^b,Q_\b\]=2\gam{}{b}{}{\a\b}Z^\a,\label{IIAsuperalgebra}\\
\[P_b,Q_\bd\]&=&-2\gam{b}{}{}{\ad\bd}Z^\ad,~~~
\[Z^b,Q_\bd\]=2\gam{}{b}{}{\ad\bd}Z^\ad,
\nonum
\eea
which is closed due to the identity (\ref{identityIIA1}),
and those undotted indices replaced by dotted spinor ones.

\win
We show that the IIA superstring action can be constructed
from the above algebra.
In this sense, we refer to the algebra as IIA superstring algebra.
The supergroup manifold is parametrized as
\bea
U=e^{{z}_aZ^a}e^{\zeta_\ad Z^\ad}e^{\zeta_\a Z^\a}e^{x^aP_a}e^{\theta^\ad Q_\ad}e^{\theta^\a Q_\a}
\eea
and the left-invariant pullback supergroup vielbeins are obtained as 
\bea
\L{i}{\a}&=&\ra_i\theta^\a,~~\L{i}{\ad}=\ra_i\theta^\ad,\\
\L{i}{a}&=&\ra_ix^a+\half(\thetadb\ga^a\ra_i\thetad)+\half(\thetab\ga^a\ra_i\theta),\\
\L{ia}{}&=&\ra_i{z}_a-\half(\thetadb\ga_a\ra_i\thetad)+\half(\thetab\ga_a\ra_i\theta),\\
\L{i\a}{}&=&\ra_i\zeta_\a+2\ra_i{z}_a(\thetab\ga^a)_\a+2\ra_ix^a(\thetab\ga_a)_\a
       +\frac{2}{3}(\thetab\ga^a\ra_i\theta)(\thetab\ga_a)_\a,\\
\L{i\ad}{}&=&\ra_i\zeta_\ad+2\ra_i{z}_a(\thetadb\ga^a)_\ad-2\ra_ix^a(\thetadb\ga_a)_\ad
       -\frac{2}{3}(\thetadb\ga^a\ra_i\thetad)(\thetadb\ga_a)_\ad,
\eea
which are invariant under the following supersymmetry transformations:
\bea
\de \theta^\a&=&\ep^\a,~~~\de \theta^\ad=\ep^\ad,\nonum\\
\de x^a&=& -\half(\epb \ga^a \theta)-\half(\epdb \ga^a \thetad),\nonum\\
\de {z}_a&=& -\half(\epb \ga^a \theta)+\half(\epdb \ga^a \thetad),
\\
\de \zeta_\a&=&-2x^a(\epb\ga_a)_\a-2{z}_a(\epb\ga^a)_\a
        +\frac{2}{3}(\epb\ga_a\theta)(\thetab\ga^a)_\a,\nonum\\
\de \zeta_\ad&=&+2x^a(\epdb\ga_a)_\ad-2{z}_a(\epdb\ga^a)_\ad
        -\frac{2}{3}(\epdb\ga_a\thetad)(\thetadb\ga^a)_\ad.\nonum
\eea
In the case where we do not denote the spinor indices explicitly,
it is always understood that they have their standard position, e.g.
$(\ga_a\theta)_\a=(\ga_a)_{\a\b}\theta^\b$,
$(\thetab\ga_a\ra_i\theta)=\theta^\a(\ga_a)_{\a\b}\ra_i\theta^\b$,
etc.

\win
We find that the IIA superstring action can be constructed as
\bea
I=\int d^2\xi\[
-\half\sqrt{-g}g^{ij}\L{i}{a}\L{j}{b}\eta_{ab}
-\half\ep^{ij}
(
\L{i}{a}\L{ja}{}
+\frac{1}{4}\L{i}{\a}\L{j\a}{}
+\frac{1}{4}\L{i}{\ad}\L{j\ad}{}
)
\],
\label{IIAaction}
\eea
where $g_{ij}$ and $\eta_{ab}$ are the worldsheet and spacetime metric,
respectively, and $g=det g_{ij}$.
The last three terms in the action constitute the manifestly supersymmetric form of the
Wess-Zumino term.
The coefficient of the Wess-Zumino term is determined
so that the action enjoys fermionic $\kappa$-symmetry,
\bea
\de_\ka\theta^\a={(1+\Ga)^\a}_\b\ka^\b,
~~~
\de_\ka\theta^\ad={(1-\Ga)^\ad}_\bd\ka^\bd,~~~
\de_\ka x^a=\half(\de_\ka\thetab\ga^a\theta)+\half(\de_\ka\thetadb\ga^a\thetad),
\eea
where $\Ga=\frac{1}{2\sqrt{-g}}\ep^{ij}\L{i}{a}\L{j}{b}\ga_{ab} $.

\win
The two-form $b$ is defined from the Wess-Zumino term in the action
as
\bea
b=-\half(\L{}{a}\wedge\L{a}{}
+\frac{1}{4}\L{}{\a}\wedge\L{\a}{}
+\frac{1}{4}\L{}{\ad}\wedge\L{\ad}{}
).
\eea
The three-form $h=db$ is calculated as
\bea
h=- \half L^a\wedge L^\a\wedge L^\b(\ga_a)_{\a\b}
   +\half L^a\wedge L^\ad\wedge L^\bd(\ga_a)_{\ad\bd}.
\eea
Note that all of the dependence on the new fermionic coordinates
has dropped out from the expression for $h$.
In fact, the anti-symmetric part of the action
is the well-known Wess-Zumino term of the IIA superstrings
up to total derivative terms,
\bea
\int\ep^{ij}b_{ij}
=\int\half\ep^{ij}
\{\ra_ix^a((\thetab\ga_a\ra_j\theta)-(\thetadb\ga_a\ra_j\thetad))
+\half(\thetadb\ga^a\ra_i\thetad)(\thetab\ga_a\ra_j\theta)
\}
\].
\eea
As a result, we conclude that the IIA superstring actin is constructed
from the IIA superstring algebra
(\ref{IIAsuperalgebra}).

\sect{IIB Superstring}\label{IIB}
\win
In this section, the IIB superalgebra is constructed as the T-dual of the IIA superalgebra.
The IIB superalgebra includes D-string charges as well as superstring charges.
We show that the IIB superstring action is constructed by means of the IIB superalgebra.

\subsection{IIB superalgebra and T-duality}\label{IIB superalgebra}
\win
In order to obtain IIB superalgebra,
we consider a T-duality transformation of the IIA superalgebra.
We denote the 9-th spacelike direction as $x^\sh$ with respect to which T-duality is performed.
The IIB superalgebra is generated by supertranslations $Q_A$, superstring $Z^A$
and D-string $\S^A$,
which can be expressed in terms of generators of the IIA superalgebra.
Since the IIB superalgebra is generated by generators with undotted spinor indices,
the chirality of the fermionic generators
with dotted spinor indices $\ad$ in the IIA superalgebra
are flipped by multiplying $\ga^\sh$,
as was done in ref.\cite{Townsend}.
We denote chirality flipped spinor indices as $\at$
and $(a,\a,\at)$ as $A$ collectively.

\win
It turns out to be easy to perform the T-duality transformation by using the Maurer-Cartan equations.
We found that the dual forms $\hat e^A$, $\hat e_A$ and $\hat e_A'$
of generators $Q_A$, $Z^A$ and $\S^A$, respectively,
can be written in terms of those of the generators of the IIA superalgebra as follows:
\bea
\begin{tabular}{rrl}
Supertranslation $Q_A$:
& $\hat e^A $&$= \Big( \hat e^a=(e^i,e_\sh),~\hat e^\a=e^\a, ~\hat e^\at =\ga^\sh e^\ad\Big)$,\\
Superstring $Z^A$:
& $\hat e_A $&$= \Big( \hat e_a=(e_i,e^\sh), ~\hat e_\a=e_\a, ~\hat e_\at=\ga^\sh e^\ad\Big)$,\\
D-string $\S^A$:
& $\hat e_A'$&$= \Big( \hat e_a'=(e_{\sh i},-e), ~{\hat e_\a}'=e_{\sh\a},
~{\hat e_\at}'=\ga^\sh e_{\sh\ad}\Big)$,
\end{tabular}
\eea
where $i$ runs, except for $\sh$.
We use a notation in which the primed dual form corresponds
to a dual form for D-string charges.
Since the dual forms $e_{\a\b}$, $e_{\a\bd}$ and $e_{\ad\bd}$ of the IIA superalgebra
are parts of the IIB D3-brane charges, we neglect them here.
The resulting Maurer-Cartan equations for the IIB superalgebra are found to be
(dropping hats)
\bea
de^a&=& -\half e^\a\wedge e^\b (\ga^a)_{\a\b}
        -\half e^\at\wedge e^\bt (\ga^a)_{\at\bt},\label{IIB1}\\
de_a&=& -\half e^\a\wedge e^\b (\ga_a)_{\a\b}
        +\half e^\at\wedge e^\bt (\ga_a)_{\at\bt},\\
de_\a&=& -e^\b\wedge e^a (\ga_a)_{\a\b}
         -e^\b\wedge e_a (\ga^a)_{\a\b},
\label{IIB3}\\
de_\at&=& +e^\bt\wedge e^a (\ga_a)_{\at\bt}
          -e^\bt\wedge e_a (\ga^a)_{\at\bt},
\label{IIB4}\\
de_a'&=&-e^\a \wedge e^\bt (\ga_a)_{\a\bt},\label{IIB5}\\
de_\a'&=& -e^\bt\wedge e^a (\ga_a)_{\a\bt}
           -e^\b\wedge e_a' (\ga^a)_{\a\b},\label{IIB6}\\
de_\at'&=& -e^\b\wedge e^a (\ga_a)_{\at\b}
          -e^\bt\wedge e_a' (\ga^a)_{\at\bt},\label{IIB7}
\eea
where the Jacobi identities are satisfied due to the well-known identity
$(\ga_a)_{\a(\b}(\ga^a)_{\ga\de)}=0$.
Here the tildes on the spinor indices of $\ga$-matrices
are not written
because the $\ga$-matrices do not see the tilded-ness of the spinors.
But if one wants to see how the identity is used,
one finds that used are the following identities:
\bea
(\ga_a)_{\a(\b}(\ga^a)_{\ga\de)}=0,~~~~
(\ga_a)_{\at(\b}(\ga^a)_{\ga\de)}=0,
\eea
and those in which the tilded spinor indices are exchanged for
the untilded spinor indices.
Note that the numbers of tildes in the identities are $0, 4, 1, 3$
and the identity with two tilded spinor indices is absent.
Since the T-dual of the Maurer-Cartan equations
(\ref{primed1}) $\sim$ (\ref{primed4})
can not be rearranged
in a covariant form,
we drop them here.
We return to this point later in sec.\ref{identities}.

\subsection{IIB superstring}
\win
We show that the IIB superstring action can be constructed from
the IIB superalgebra obtained in sec.\ref{IIB superalgebra}.
We start with writing down the IIB superalgebra:
\bea
\{Q_\a,Q_\b\}&=&\gam{}{a}{}{\a\b}P_a+ \gam{a}{}{}{\a\b}Z^a,\nonum\\
\{Q_\at,Q_\bt\}&=&\gam{}{a}{}{\at\bt}P_a- \gam{a}{}{}{\at\bt}Z^a,\nonum\\ 
\{Q_\a,Q_\bt\}&=&\gam{a}{}{}{\a\bt}\S^a,\nonum\\
\[P_a,Q_\b\]&=&2\gam{a}{}{}{\a\b}Z^\a + 2\gam{a}{}{}{\at\b}\S^\at,\\
\[P_a,Q_\bt\]&=&-2\gam{a}{}{}{\at\bt}Z^\at + 2\gam{a}{}{}{\a\bt}\S^\a,\nonum\\
\[Z^a,Q_\b\]&=&2\gam{}{a}{}{\a\b}Z^\a,~~~
\[Z^a,Q_\bt\]=2\gam{}{a}{}{\at\bt}Z^\at,\nonum\\
\[\S^a,Q_\b\]&=&2\gam{}{a}{}{\a\b}\S^\a,~~~
\[\S^a,Q_\bt\]=2\gam{}{a}{}{\at\bt}\S^\at.\nonum
\eea

\win
Parametrizing the supergroup manifold by
\bea
U=e^{z_aZ^a}e^{\zeta_\at Z^\at}e^{\zeta_\a Z^\a}
e^{y_a\S^a}e^{\phi_\at\S^\at}e^{\phi_\a\S^\a}
e^{x^aP_a}e^{\theta^\at Q_\at}e^{\theta^\a Q_\a},
\eea
we obtain the pullback vielbeins of the left-invariant supergroup as follows:
\bea
\L{i}{\a}&=&\ra_i\theta^\a,~~~\L{i}{\at}=\ra_i\theta^\at,
\label{IIBL1}\\
\L{i}{a}&=&\ra_ix^a
         +\half(\thetab\ga^a\ra_i\theta)
         +\half(\thetatb\ga^a\ra_i\thetat),\\
\L{ia}{}&=&\ra_iz_a
         +\half(\thetab\ga^a\ra_i\theta)
         -\half(\thetatb\ga^a\ra_i\thetat),\\
\L{i\a}{}&=&\ra_i\zeta_\a
           +2\ra_iz_a(\thetab\ga^a)_\a
           +2\ra_ix^a(\thetab\ga_a)_\a
           +\frac{2}{3}(\thetab\ga^a\ra_i\theta)(\thetab\ga_a)_\a,\\
\L{i\at}{}&=&\ra_i\zeta_\at
           +2\ra_iz_a(\thetatb\ga^a)_\at
           -2\ra_ix^a(\thetatb\ga_a)_\at
           -\frac{2}{3}(\thetatb\ga^a\ra_i\thetat)(\thetatb\ga_a)_\at,\\
{\L{ia}{}}'&=&\ra_iy_a
             +(\thetatb\ga_a\ra_i\theta),\label{IIBL6}\\
{\L{i\a}{}}'&=&\ra_i\phi_\a
             +2\ra_ix^a(\thetatb\ga_a)_\a
             +2\ra_iy_a(\thetab\ga^a)_\a
             +(\thetab\ga^a\ra_i\theta)(\thetatb\ga_a)_\a
             +\frac{1}{3}(\thetatb\ga^a\ra_i\thetat)(\thetatb\ga_a)_\a,
\label{IIBL7}\\
{\L{i\at}{}}'&=&\ra_i\phi_\at
             +2\ra_ix^a(\thetab\ga_a)_\at
             +2\ra_iy_a(\thetatb\ga^a)_\at
             +(\thetatb\ga^a\ra_i\theta)(\thetatb\ga_a)_\at
             +\frac{1}{3}(\thetab\ga^a\ra_i\theta)(\thetab\ga_a)_\at.
\label{IIBL8}
\eea
The supersymmetry transformations are found to be
\bea
\de \theta^\a &=& \ep^\a ,~~~\de \theta^\at=\ep^\at,
\label{IIBsusy1}\\
\de x^a&=& -\half (\epb\ga^a\theta)
           -\half (\eptb \ga^a\thetat),\\
\de z_a&=& -\half (\epb\ga^a\theta)
           +\half (\eptb \ga^a\thetat),\\
\de \zeta_\a&=& -2z_a(\epb\ga^a)_\a -2x^a(\epb\ga_a)_\a
             +\frac{2}{3} (\epb \ga_a\theta)(\thetab\ga^a)_\a,\\
\de \zeta_\at&=& -2z_a(\eptb\ga^a)_\at +2x^a(\eptb\ga_a)_\at
             -\frac{2}{3} (\eptb \ga_a\thetat)(\thetatb\ga^a)_\at,\\
\de y_a&=&-(\eptb \ga_a\theta),
\label{IIBsusy6}\\
\de \phi_\a&=& -2x^a(\eptb \ga_a)_\a
               -2y_a(\epb\ga^a)_\a
               +(\eptb\ga^a\theta)(\thetab\ga_a)_\a
               +\frac{1}{3}(\eptb\ga^a\thetat)(\thetatb\ga_a)_\a,\\
\de \phi_\at&=& -2x^a(\epb \ga_a)_\at
               -2y_a(\eptb\ga^a)_\at
               +(\eptb\ga^a\thetat)(\thetab\ga_a)_\at
               +\frac{1}{3} (\epb\ga^a\theta)(\thetab\ga_a)_\at.
\eea
We find that the IIB superstring action is constructed as
\BE
I=\int d^2\xi \[
-\half\sqrt{-g}g^{ij}L_i^{~a}L_{j}^{~b}\eta_{ab}
-\half \ep^{ij} 
(L_i^{a}L_{ja}+\frac{1}{4}L_i^{\a}L_{j\a}+\frac{1}{4}L_i^{~\at}L_{j\at})
\],
\EE
where $g_{ij}$ and $\eta_{ab}$ are the worldsheet and the spacetime metric, respectively,
and $g=det g_{ij}$.
The last three terms in the action constitute a manifestly supersymmetric
form of the Wess-Zumino term.
The two-form $b$ is defined in terms of the Wess-Zumino term in the action
as
\bea
b=-\half(\L{}{a}\wedge\L{a}{}
+\frac{1}{4}\L{}{\a}\wedge\L{\a}{}
+\frac{1}{4}\L{}{\at}\wedge\L{\at}{}
),
\eea
and the-three form $h=db$ is calculated as
\bea
h=- \half L^a\wedge L^\a\wedge L^\b(\ga_a)_{\a\b}
   +\half L^a\wedge L^\at\wedge L^\bt(\ga_a)_{\at\bt}.
\label{IIBh}
\eea
Note that all of the dependence on the new fermionic coordinates
has dropped out from the expression of $h$.
In fact, the anti-symmetric part of the action
is the well-known Wess-Zumino term of the IIB superstring
up to total derivative terms,
\bea
\int\ep^{ij}b_{ij}
=\int\half \ep^{ij}
\[\ra_i x^a((\thetab\ga_a\ra_j\theta)-(\thetatb\ga_a\ra_j\thetat))
+\half(\thetatb\ga_a\ra_i\thetat)(\thetab\ga^a\ra_j\theta)
\].
\label{b_{ij}}
\eea
The coefficient of the Wess-Zumino term is chosen so that
the action is invariant under the $\kappa$-symmetry transformations
\bea
\de_\ka\theta^\a =(1+\Ga)^{\a}_{~\b}\ka^\b,~~~
\de_\ka \theta^\at=(1-\Ga)^{\at}_{~\bt}\ka^\bt,~~~
\de_\ka x^a = -\half(\thetatb\ga^a\de_\ka \thetat)
          -\half(\thetab\ga^a\de_\ka \theta),
\eea
where $\Ga=\frac{1}{2\sqrt{-g}}\ep^{ij}L_i^{~a}L_j^{~b}\ga_{ab}$.

\win
In this way, we construct the IIB superstring action
from the IIB superalgebra.
Note that by constructing the IIB superstring action,
the D-string charges, $\S^A$, play no role.
This is consistent with $\S^A$ representing D-string charges.
In this sense, we refer to the algebra corresponding to the Maurer-Cartan
equations (\ref{IIB1}) $\sim$ (\ref{IIB4}) as the IIB superstring algebra.

\section{IIB D-string}
\win
The D-string action characterized by left-invariant vielbeins correspond to D-string charges
can not be constructed by means of the IIB superalgebra.
We show that the Wess-Zumino term and the modified 2-form field strength,
$\CF=F-b$, where $F$ is the 2-form field strength
and $b$ is the pullback
to the worldsheet of the $R\otimes R$ two-form gauge potential,
can be constructed in terms of the left-invariant vielbeins corresponding to D-string charges
if we start with a superalgebra obtained by {\it modifying} the IIB superalgebra.

\win
We start with the {\it modified} IIB superalgebra: (\ref{IIB1}) $\sim$ (\ref{IIB5})
and
\bea
de_\a'&=&-e^\b\wedge e^a(\ga_a)_{\a\b}-e^\bt\wedge e_a'(\ga^a)_{\a\bt},
\label{modIIB1}\\
de_\at'&=&-e^\bt\wedge e^a(\ga_a)_{\at\bt}-e^\b\wedge e_a'(\ga^a)_{\at\b},
\label{modIIB2}
\eea
which is closed due to the well-known identity
$
(\ga_a)_{\a(\b}(\ga^a)_{\ga\de)}=0.
$
As in sec.4.1, if one writes tildes on the spinor indices of the $\ga$-matrices,
one obtains the following identities:
\bea
(\ga_a)_{\a(\b}(\ga^a)_{\ga\de)}=0,~~~
(\ga_a)_{\a(\b}(\ga^a)_{\gat\det)}=0,
\eea
and those in which the tilded spinor indices are exchanged for the untilded spinor indices.
The numbers of tildes in these identities are $0,4,2$
in contrast to the identities in the IIB superalgebra,
where the numbers of tildes were $0,4,1,3$.
The differnce in the number of tildes
tells us that the modified IIB superalgebra is not related to
the IIA superlagebra by the T-duality transformation,
as shown in sec.6.
The modification is simply to interchange $\S^\a$ with $\S^\at$.
This is trivial from the IIB perspective,
because the generators, $\S^\a$ and $\S^\at$, are not distinguished inherently.
From the IIA perspective, however, this is nontrivial,
since a spinor index with a tilde corresponds to
one with a different chirality under T-duality.

\win
The left-invariant vielbeins are found to be (\ref{IIBL1})$\sim$(\ref{IIBL6}) and
\bea
{\L{i\a}{}}'&=&\ra_i\phi_\a
             +2\ra_ix^a(\thetab\ga_a)_\a
             +2\ra_iy_a(\thetatb\ga^a)_\a
+(\thetatb\ga^a\ra_i\theta)(\thetatb\ga_a)_\a
         +\frac{1}{3}(\thetab\ga^a\ra_i\theta)(\thetab\ga_a)_\a,\\
{\L{i\at}{}}'&=&\ra_i\phi_\at
             +2\ra_ix^a(\thetatb\ga_a)_\at
             +2\ra_iy_a(\thetab\ga^a)_\at
+(\thetab\ga^a\ra_i\theta)(\thetatb\ga_a)_\at
        +\frac{1}{3}(\thetatb\ga^a\ra_i\thetat)(\thetatb\ga_a)_\at.
\eea
The supersymmetry transformations are obtained as (\ref{IIBsusy1})$\sim$(\ref{IIBsusy6}) and
\bea
\de \phi_\a&=& -2x^a(\epb \ga_a)_\a
               -2y_a(\eptb\ga^a)_\a
               +(\eptb\ga^a\thetat)(\thetab\ga_a)_\a
               +\frac{1}{3} (\epb\ga^a\theta)(\thetab\ga_a)_\a,\\
\de \phi_\at&=& -2x^a(\eptb \ga_a)_\at
               -2y_a(\epb\ga^a)_\at
               +(\eptb\ga^a\theta)(\thetab\ga_a)_\at
               +\frac{1}{3}(\eptb\ga^a\thetat)(\thetatb\ga_a)_\at.
\eea
The two-form $b$ is defined as
\bea
b=-\frac{1}{4}(
   \L{}{\a}\wedge {\L{\a}{}}'
  -\L{}{\at}\wedge {\L{\at}{}}'
).
\eea
The primed vielbeins correspond to the D-string charges
$\S^A$,
and the action constructed with
this Wess-Zumino term can be regarded as a ``gauge-fixed'' D-string action.
In fact, the three-form $h=db$ turns out to be
(\ref{IIBh}) obtained for the IIB superstring.
All of the dependence on the new fermionic coordinates
has dropped out from the expression of $h$.

\win
Can the total (gauge-unfixed) D-string action be constructed?
To this end, we must first determine a superinvariant modified 2-form field strength,
$\CF=dA-b$, where $A$ is a U(1) gauge field on the world-sheet.
The $b$ is the conventional 2-form which is read off from the integrand of the r.h.s
of (\ref{b_{ij}}).
If we obtain the two-form $\CF$, the D-string antion can be constructed
as in ref.\cite{Schwarz} or in ref.\cite{scale-invariant2}.
We observe that
\bea
\half\ep^{ij}(\L{i}{\a}{\L{j\a}{}}'-\L{i}{\at}{\L{j\at}{}}')
=-2\ep^{ij}(F_{ij}-b_{ij}),
\eea
where we introduce $F_{ij}$ as
\bea
\ep^{ij}F_{ij}=-\half\ep^{ij}
\[
\ra_iy_a\ra_j(\thetatb\ga^a\theta)
+\half\ra_i\phi_\a\ra_j\theta^\a
+\half\ra_i\phi_\at\ra_j\theta^\at
\].
\label{F}
\eea
The r.h.s. of (\ref{F}) is a total derivative term,
and then we can regard $F_{ij}$
as the field strength of a U(1) gauge field $A_i$,
\bea
A_i=-\half\[y_a\ra_i(\thetatb\ga^a\theta)
           +\half\phi_\a\ra_i\theta^\a
           -\half\phi_\at\ra_i\theta^\at\].
\eea
Note that this is parametrized by D-string coordinates associated to
D-string charges $\S^A$.
The supersymmetry transformation is found to be
\bea
\de A_i= -\half\[
-x^a(\epb\ga_a\ra_i\theta)
+\frac{1}{6}(\epb\ga_a\theta)(\thetab\ga^a\ra_i \theta)
-(\theta\rightarrow\thetat,~\ep\rightarrow\ept)\]
-\frac{1}{4}\ra_i((\eptb\ga_a\theta)(\thetab\ga^a\thetat)).
\eea
In order to see the relation to the well-known supersymmetry
transformation of the U(1) gauge field in ref.\cite{Schwarz},
we consider a U(1) gauge transformation and obtain an alternative form,
\bea
A_i=\half \[\ra_iy_a(\thetatb\ga^a\theta)
           +\half\ra_i\phi_\a\theta^\a
           -\half\ra_i\phi_\at\theta^\at\].
\eea
This transforms under the supersymmetry transformation as
\bea
\de A_i&=&-\half\[
\ra_ix^a(\epb\ga_a\theta)
+\frac{1}{6}(\epb\ga_a\theta)(\thetab\ga^a\ra_i \theta)
-(\theta\rightarrow\thetat,~\ep\rightarrow\ept)\]
-\frac{1}{4}(\ra_i\phi_\a\ep^\a -\ra_i\phi_\at\ep^\at ),
\label{deA}
\eea
which is similar to the well-known form~\cite{Schwarz} except for the last two terms.
These two terms are the total derivative terms,
and the supersymmetry transformation $\de F_{ij}$
is nothing but the one obtained there.
We conclude that the U(1) gauge field on the worldsheet can be
constructed in this way.
It is interesting that the U(1) gauge field is constructed explicitly
in terms of the D-string charges.
Using the modified field strength $\CF$,
one constructs the D-string action, as in ref.\cite{Schwarz} or in \cite{scale-invariant2}.
In this sence, we refer to the superalgebra corresponding to
the Maurer-Cartan equations (\ref{IIB1}), (\ref{IIB5}), (\ref{modIIB1}) and (\ref{modIIB2})
as D-string superalgebra.
Hence, the modified IIB superalgebra describes
IIB superstrings and D-strings on an equal footing.

\win
We now comment on the existence of a U(1) gauge field in type-II superstrings.
We observe that the modified field strength for IIB superstrings is constructed
by observing that
\bea
\ep^{ij}( \L{i}{a}\L{ja}{}+\frac{1}{4}\L{i}{\a}\L{j\a}{}+\frac{1}{4} \L{i}{\at}\L{j\at}{})
=-2\ep^{ij}(F_{ij}-b_{ij}),
\eea
where $F_{ij}$ is introduced as
\bea
\ep^{ij}F_{ij}=-\half\ep^{ij}\[
\ra_ix^a\ra_iz_a 
+\frac{1}{4}\ra_i\zeta_\a\ra_j\theta^\a
+\frac{1}{4}\ra_i\zeta_\at\ra_j\theta^\at
\],
\eea
and the $b_{ij}$ is the conventional 2-form defined by
the integrand of the r.h.s. of (\ref{b_{ij}}).
Thus the U(1) gauge field can be defined as
\bea
A_i=\half\[
z_a\ra_ix^a 
-\frac{1}{4}\zeta_\a\ra_i\theta^\a
-\frac{1}{4}\zeta_\at\ra_i\theta^\at
\]
\eea
and is parametrized by coordinates corresponding to IIB superstring charges.
The supersymmetry transformation is found to be
\bea
\de A_i&=&-\half\[
\ra_ix^a(\epb\ga_a\theta)
+\frac{1}{6}(\epb\ga_a\theta)(\thetab\ga^a\ra_i\theta)
-\half\ra_i(x^a(\epb\ga_a\theta))
-(\theta\rightarrow\thetat,~\ep\rightarrow\ept)\],
\eea
which is identical to (\ref{deA}) up to a total derivative term.
In this way, we can construct a U(1) gauge field for IIB superstrings.
For IIA superstrings, one can construct a U(1) gauge field parametrized by
coordinates corresponding to IIA superstring charges in a similar way.
The fact that a U(1) gauge field can be constructed in type-II superstring theories
is consistent with the space-time scale-invariant formulation
of $p$-branes and type-II superstrings~\cite{scale-invariant1,scale-invariant2}.
For $p$-brane theories, the same procedure presented above will produce
not only the supersymmetry transformation of the $p$-form gauge field,
but also the explicite form of the $p$-form gauge field
in terms of the $p$-brane charges.

\section{T-duality and Identities}
\label{identities}
\win
The modified IIB superalgebra will not be related by the T-duality to the IIA superalgebra.
This is seen by recognizing the fact that identities characterizing the IIA superalgebra
can not be written in a covariant form after the T-duality transformation.

\win
For completeness, we start with identities in the M-algebra and reduce to ones in the IIA superalgebra.
Then, performing the T-duality transformation, we determine whether the resulting identities
are rearranged in a covariant form or not.

\win
One of the two identities characterizing the M-algebra is (\ref{identityM1})
\bea
(\ga_\mu)_{(\a\b}(\ga^{\mu\nu})_{\ga\de)}=0.
\eea
We reduce this identity to ones of the IIA superalgebra as follows.
For $\nu=b\not=\na$, one obtains the identities
\bea
&&
(1)_{\ad(\b}(\ga^a)_{\ga\de)}+({\ga^a}_b)_{\ad(\b}(\ga^b)_{\ga\de)}=0\label{identity1},
\\&&
(1)_{\a(\bd}(\ga^a)_{\gad\ded)}+({\ga^a}_b)_{\a(\bd}(\ga^b)_{\gad\ded)}=0,\label{identity2}
\eea
which are the characteristic identities in the presence of D0 and D2-branes.
As for $\nu=\na$, the well-known identities
\bea
(\ga_a)_{\a(\b}(\ga^a)_{\ga\de)}=0,
~~~
(\ga_a)_{\ad(\bd}(\ga^a)_{\gad\ded)}=0\label{identity3}
\eea
are obtained.
We next consider the T-duality transformation.
By the procedure explained in sec.\ref{IIB superalgebra},
we obtain the T-dual of the identity (\ref{identity1}) for $a=i\not=\sh$
\bea
(\ga^\sh)_{\at(\b}(\ga^i)_{\ga\de)}-(\ga^i)_{\at(\b}(\ga^\sh)_{\ga\de)}
+({\ga^{\sh i}}_{j})_{\at(\b}(\ga^j)_{\ga\de)}=0,
\eea
which is rewritten in a covariant form as
\bea
(\ga^{\[a})_{\at(\b}(\ga^{b\]})_{\ga\de)}
+\half({\ga^{ab}}_c)_{\at(\b}(\ga^c)_{\ga\de)}=0.
\label{identityIIB1}
\eea
This is a characteristic identity in the presence of
D3-branes.
In turn, for $a=\sh$, one finds the identity
\bea
(\ga_a)_{\at(\b}(\ga^a)_{\ga\de)}=0,\label{identityIIB2}
\eea
which is used in satisfying the Jacobi identities for the IIB superalgebra.
The T-dual of the identity (\ref{identity2}) is found to be identities
(\ref{identityIIB1}) and (\ref{identityIIB2}) with exchanging the tilded spinor indices
for the untilded spinor indices.
Identities (\ref{identity3}) are transformed into
\bea
(\ga_a)_{\a(\b}(\ga^a)_{\ga\de)}=0,~~~
(\ga_a)_{\at(\bt}(\ga^a)_{\gat\det)}=0,
\eea
respectively.

\win
The second identity of the M-algebra is (\ref{identityM2})
\bea
(\ga_\mu)_{(\a\b}(\ga^\mu)_{\ga\de)}
+\frac{1}{10}(\ga_{\mu\nu})_{(\a\b}(\ga^{\mu\nu})_{\ga\de)}=0,
\eea
which reduces to the identity in the IIA superalgebra,
\bea
(\phib\phid)(\phib\phid)+\frac{2}{5}(\phib\ga_a\phi)(\phidb\ga^a\phid)
+\frac{1}{10}(\phib\ga_{ab}\phid)(\phib\ga^{ab}\phid)=0,
\eea
where $\phi$ and $\phid$ are Grassmann-even spinors with opposite chirality each other.
This is characteristic identity for the Maurer-Cartan equations
(\ref{primed2}) and (\ref{primed3}) for primed dual forms. 
Under the T-duality transformation,
this transforms into an identity
which is not rewritten in a covariant form,
\bea
(\phib\ga^\sh\phit)(\phib\ga^\sh\phit)
+\frac{2}{5}(\phib\ga_i\phi)(\phitb\ga^i\phit)
-\frac{2}{5}(\phib\ga^\sh\phi)(\phitb\ga^\sh\phit)
+\frac{1}{10}(\phib\ga_{ij\sh}\phit)(\phib\ga^{ij\sh}\phit)
+\frac{1}{5}(\phib\ga_i\phit)(\phib\ga^i\phit)=0.
\eea
This was the reason why the ``superstring'' charges in the IIA superalgebra
are discarded in performing the T-duality transformation
in sec.\ref{IIB superalgebra}.

\win
Let us consider, conversely, the identity in the modified IIB superalgebra,
\bea
(\ga_a)_{\a(\b}(\ga^a)_{\gat\det)}=0.\label{identitymodifiedIIB}
\eea
This is transformed into an identity,
\bea
(\ga_i)_{\a\b}(\ga^i)_{\gad\ded}
-(\ga_\sh)_{\a\b}(\ga^\sh)_{\gad\ded}
+(\ga_{i\sh})_{\a\gad}(\ga^{i\sh})_{\ded\b}
+(\ga_{i\sh})_{\a\ded}(\ga^{i\sh})_{\b\gad}
+(1)_{\a\gad}(1)_{\b\ded}
+(1)_{\a\ded}(1)_{\b\gad}
=0,
\eea
which is not rewritten in a covariant form.

\win
From these observation, the modified IIB superalgebra is not rewritten as
a T-dual of the IIA superalgebra.
In the next section,
we try to relate the IIA superalgebra and the modified IIB superalgebra.

\sect{Unification of Type-II Superalgebras}
\win
We encountered two IIB superalgebras.
One is the IIB superalgebra, which is the T-dual of the IIA superalgebra;
the other is the modified IIB superalgebra,
which describes
the IIB superstring and D-string on an equal footing.
The IIA superalgebra is related to the IIB superalgebra by the T-duality,
but not to the modified IIB superalgebra.

\win
As a trial to relate the IIA superalgebra to
the modified IIB superalgebra,
we examine a unification in a
$(10+1)$-dimensional $N=2$ superalgebra
of the modified IIB superalgebra and the M-algebra
(hence, the IIA superalgebra).
This is motivated by the fact that
the identity in $N=2$ $D=10+1$,
\bea
(\ga_\mu)_{(\a\b}(\ga^{\mu\nu})_{\gat\det)}=0,
\eea
is projected into the identity (\ref{identitymodifiedIIB})
in the modified IIB superalgebra.

\win
We first present the relevant part of $(10+1)$-dimensional $N=2$ superalgebra,
and then consider the relations to the M-algebra and the modified IIB superalgebra.
We begin with the $N=2$ $(10+1)$-dimensional superalgebra
generated by bosonic charges:
\bea
de^\mu&=& -\half e^\a\wedge e^\b (\ga^\mu)_{\a\b}
        -\half e^\at\wedge e^\bt (\ga^\mu)_{\at\bt},\\
de_\mu&=& -\half e^\a\wedge e^\b (\ga_\mu)_{\a\b}
        +\half e^\at\wedge e^\bt (\ga_\mu)_{\at\bt},\\
de_{\mu \nu}&=& -\half e^\a\wedge e^\b (\ga_{\mu \nu})_{\a\b}
           -\half e^\at\wedge e^\bt (\ga_{\mu\nu})_{\at\bt},\\
de^{\mu\nu}&=& -\half e^\a\wedge e^\b (\ga^{\mu\nu})_{\a\b}
           +\half e^\at\wedge e^\bt (\ga^{\mu\nu})_{\at\bt},\\
de_{\mu\nu}'&=& - e^\a\wedge e^\bt (\ga_{\mu\nu})_{\a\bt},\\
de_{\mu}'&=& - e^\a\wedge e^\bt (\ga_{\mu})_{\a\bt}.
\eea
In addition to these equations we consider the following equations:
\bea
de_\a&=& -e^\b \wedge e^\mu (\ga_\mu)_{\a\b}
         +(1-\la) e^\b\wedge e_\mu(\ga^\mu)_{\a\b}
        -\frac{1}{10}e^\b\wedge e_{\mu\nu}(\ga^{\mu\nu})_{\a\b}
\nonum\\
&&         +\frac{1}{10}(1-\la)e^\b\wedge e^{\mu\nu}(\ga_{\mu\nu})_{\a\b}
       -\frac{1}{10}(2-\la)e^\bt\wedge e_{\mu\nu}' (\ga^{\mu\nu})_{\a\bt}
        -(2-\la)e^\bt\wedge e_\mu' (\ga^\mu)_{\a\bt},
\label{1ofN=2}\\
de_\at&=& -e^\bt \wedge e^\mu (\ga_\mu)_{\at\bt}
          +e^\bt \wedge e_\mu(\ga^\mu)_{\at\bt}
        +\frac{1}{10}e^\bt \wedge e_{\mu\nu}(\ga^{\mu\nu})_{\at\bt}
        -\frac{1}{10}e^\bt \wedge e^{\mu\nu}(\ga_{\mu\nu})_{\at\bt},
\label{2ofN=2}\\
de_{\mu\a}&=& -e^\b \wedge e^\nu(\ga_{\mu\nu})_{\a\b}
            -e^\b \wedge e_{\mu\nu}(\ga^{\nu})_{\a\b}
            -e^\bt \wedge {e_{\mu\nu}}'(\ga^{\nu})_{\a\bt}
            -e^\bt \wedge {e_\nu}'({\ga_{\mu}}^\nu)_{\a\bt}\nonum\\
   &&       -e^\bt \wedge e^\nu(\ga_{\mu\nu})_{\a\bt}
            -e^\bt \wedge e_{\mu\nu}(\ga^{\nu})_{\a\bt}
            -e^\b \wedge {e_{\mu\nu}}'(\ga^{\nu})_{\a\b}
            -e^\b \wedge {e_\nu}'({\ga_{\mu}}^\nu)_{\a\b},
\label{3ofN=2}\\
de_{\mu\at}&=& -e^\bt \wedge e^\nu(\ga_{\mu\nu})_{\at\bt}
            -e^\bt \wedge e_{\mu\nu}(\ga^{\nu})_{\at\bt}
            -e^\b \wedge {e_{\mu\nu}}'(\ga^{\nu})_{\at\b}
            -e^\b \wedge {e_\nu}'({\ga_{\mu}}^\nu)_{\at\b}\nonum\\
   &&       -e^\b \wedge e^\nu(\ga_{\mu\nu})_{\at\b}
            -e^\b \wedge e_{\mu\nu}(\ga^{\nu})_{\at\b}
            -e^\bt \wedge {e_{\mu\nu}}'(\ga^{\nu})_{\at\bt}
            -e^\bt \wedge {e_\nu}'({\ga_{\mu}}^\nu)_{\at\bt},
\label{4ofN=2}\\
de_{\a\b}&=& \half e^\mu\wedge e^\nu (\ga_{\mu\nu})_{\a\b}
            +\half e^\mu\wedge e_{\mu\nu} (\ga^\nu)_{\a\b}
            -2 e^\ga\wedge e_{\mu\a} (\ga^\mu)_{\b\ga}
            -\frac{1}{4} e^\ga\wedge e_{\mu\ga} (\ga^\mu)_{\a\b}\nonum\\
  &&        -2 e^\gat\wedge e_{\mu\a} (\ga^\mu)_{\b\gat}
            -\frac{1}{4} e^\gat\wedge e_{\mu\gat} (\ga^\mu)_{\a\b}
            +\half {e_\mu}'\wedge e_{\nu}' (\ga^{\mu\nu})_{\a\b}
            + e^\mu\wedge e_{\nu}' ({\ga_\mu}^\nu)_{\a\b}   \nonum\\
 &&         +\half e^\mu\wedge e_{\mu\nu}' (\ga^\nu)_{\a\b}
            +\half e_\mu'\wedge e_{\nu\la}' (\ga^\la)_{\a\b}\eta^{\mu\nu}
            +\half e_\mu'\wedge e_{\nu\la} (\ga^\la)_{\a\b}\eta^{\mu\nu}.
\label{5ofN=2}
\eea
The Jacobi identity for the dual form $de_{\mu\a}$
is satisfied such that the exterior derivatives of
the first and second lines in (\ref{3ofN=2})
vanish separately.
However, the Jacobi identity for the dual form $de_{\a\b}$
requires both lines in (\ref{3ofN=2}).

\win
Let us consider the relation to the M-algebra.
We discard the dual forms with spinor indices with a tilde,
since we need only $N=1$ supersymmetry.
In addition, we discard the primed dual forms.
These procedures cause the following to occur:
eqns.(\ref{2ofN=2}) and (\ref{4ofN=2}) decouple from our analysis;
eqn.(\ref{1ofN=2}), with identifying $e^{\mu\nu}$ with $e_{\mu\nu}$
turns out to be (\ref{3inMalgebra});
eqns.(\ref{3ofN=2}) and (\ref{5ofN=2}) result in
(\ref{5inMalgebra}) and (\ref{6inMalgebra}), respectively.
We thus find that the N=2 superalgebra includes the M-algebra in this way.

\win
We next consider the relation to the modified IIB superalgebra.
We project the spinor indices to ones with the same chirality
and perform a dimensional reduction of the 11-th dimension $x^\na$.
The dual forms $e_{\na a}$, $e^{\na a}$ and $e_{\na a}'$ are
identified with $e^a$, $e_a$ and $e_a'$, respectively.
We find that if we set $\la=2$,
eqn.(\ref{1ofN=2}) turns to (\ref{IIB3})
after a trivial overall scaling.
Eqn.(\ref{2ofN=2}) is found to become (\ref{IIB4}).
Eqns.(\ref{3ofN=2}) and (\ref{4ofN=2}) with $\mu=\na$,
after a trivial overall rescaling,
turn out to be
\bea
de_\a'&=& -e^\bt\wedge e^a (\ga_a)_{\a\bt}
           -e^\b\wedge e_a' (\ga^a)_{\a\b}
           -e^\b\wedge e^a(\ga_a)_{\a\b}
           -e^\bt\wedge e_a'(\ga^a)_{\a\bt},
\\
de_\at'&=& -e^\b\wedge e^a (\ga_a)_{\at\b}
          -e^\bt\wedge e_a' (\ga^a)_{\at\bt}
          -e^\bt\wedge e^a(\ga_a)_{\at\bt}
          -e^\b\wedge e_a'(\ga^a)_{\at\b}.
\eea
Note that the r.h.s. of these equations are
constructed by adding the r.h.s. of (\ref{IIB6}) and (\ref{IIB7})
of the IIB superalgebra
and the r.h.s. of (\ref{modIIB1}) and (\ref{modIIB2})
of the modified IIB superalgebra.
The rest of the equations will be a part of D3-brane charges.

\win
In summary, we find that
the $N=2$ superalgebra includes the M-algebra
and the free parameter $\lambda$ in the M-algebra
has to be $2$ in order for the IIB superstring algebra
to be included.
These imply that
(\ref{modIIB1}) and (\ref{modIIB2}) for the D-string superalgebra
naturally arise as well as (\ref{IIB6}) and (\ref{IIB7}) of the IIB superalgebra.
Note that
the commutators of the D-string superalgebra,
which are not related to the IIA superalgebra
by the T-duality,
emerge by considering the $(10+1)$-dimensional $N=2$ superalgebra. 
However, D-strings do not correspond to the superalgebra
obtained from the $N=2$ superalgebra,
since the pullback vielbeins, ${\L{i\a}{}}'$ and ${\L{i\at}{}}'$,
contain the r.h.s. of (\ref{IIBL7}) and (\ref{IIBL8}).
It is interesting to seek
a unified superalgebra from which one can construct
IIA superstrings, IIB superstrings and D-strings.

\sect{Summary and Discussions}
\win
We have presented a set of new spacetime superalgebras:
the IIA superstring superalgebra, the IIB superstring
superalgebra and the D-string superalgebra.
Using the new superalgebras,
we have shown that Siegel's formulation generalizes
to type-II superstrings and D-strings.
Namely,
we have constructed supercurrents on the supergroup manifolds
corresponding to the superalgebras.
We then wrote down the Wess-Zumino terms,
which are second order in the supercurrents.
The modified 2-form field strength for D-strings
was identified with a second-order expression of the supercurrents.
From this expression,
the U(1) gauge field on the worldsheet of D-strings
was obtained explicitly in terms of coordinates corresponding
to D-string charges,
including the new fermionic charges.

\win
We succeeded in constructing
the pullback to the D-string worldsheet of the $R\otimes R$ 2-form potential
from the modified IIB superalgebra in sec.5.
We now comment on the relation to D-strings of the IIB superalgebra,
which was obtained in sec.4 by a T-duality transformation
of the IIA superalgebra.
We observe that
using the pullbacks of left-invariant vielbeins
(\ref{IIBL1}) $\sim$ (\ref{IIBL8})
corresponding to the IIB superalgebra,
the second-order expression,
$
-\half( L^a\wedge L_a'+\frac{1}{4}L^\a\wedge L_\a' +\frac{1}{4}L^\at\wedge L_\at')
$,
is calculated to be, up to total derivative terms,
\bea
-\half\Big( dx^a\wedge((\thetatb\ga_ad\theta)+(\thetab\ga_a d\thetat))
+\frac{1}{3}(\thetab\ga^ad\theta)\wedge(\thetab\ga_ad\thetat)
+\frac{1}{3}(\thetatb\ga^ad\thetat)\wedge(\thetatb\ga_ad\theta)
\Big),
\eea
which corresponds to the pullback to the D-string worldsheet of the $NS\otimes NS$ 2-form
potential.
Thus we find that interchanging new fermionic generators $\S^\a$ and $\S^\at$
results in exchanging $R\otimes R$ gauge potential for $NS\otimes NS$ one,
since the modification was simply interchanging 
$\S^\a$ and $\S^\at$.

\win
In turn, 
the pullback to the F-string worldsheet of the $R\otimes R$ 2-form
potential can be obtained from
the pullback to the D-string worldsheet of the $R\otimes R$ 2-form
potential.
Using the resulting expressions, we can construct $(p,q)$-superstring actions
in a manifestly supersymmetric form.
We hope to report on this issue in the future~\cite{sakagu}.
In addition, it is interesting to see whether
the formulation can be generalized to the other type-II-branes:
NS5-branes and D $p$-branes
(p$=$odd for the IIB superstring theory
and p$=$even for the IIA superstring theory).
Especially, now that we have the IIA superalgebra, including
the D2-brane charges, the generalization to the D2-branes
has to be examined.
We leave this to the future.

\win
We found a modified IIB superalgebra which includes the IIB superstring
and D-string superalgebras as subalgebras,
and describes IIB superstrings and D-strings
on an equal footing.
However, this
is not related by the T-duality transformation
to the IIA superalgebra derived from the M-algebra.
In order to relate these superalgebras,
we considered a unification
in a $(10+1)$-dimensional $N=2$ superalgebra.
The unification implies that
the free parameter in the M-algebra is fixed as $\la=2$.
By considering the unification,
the commutation relations of D-strings,
which was not obtained by
the T-duality transformation of the IIA superalgebra,
are found to emerge.
However, unnecessary relations are also generated
in addition to
the preferred D-string superalgebra,
and the obtained superalgebra does not correspond to D-strings.
It is interesting to consider the other unification which
unifies the M-algebra and the modified IIB superalgebra.
\vs{5}

{\bf Acknowledgments}

\win
The author would like to express his gratitude to Profs. N. Ishibashi and
K. Hamada.
This research was supported in part by JSPS Research Fellowships for Young Scientists.



\end{document}